\newcommand{\be}{\begin{equation}}\newcommand{\ee}{\end{equation}}
\newcommand{\bea}{\begin{eqnarray}}\newcommand{\eea}{\end{eqnarray}}
\newcommand{\nn}{\nonumber \\}
\newcommand{\hK}{{\hat K}}
\newcommand{\hD}{{\hat D}}

\newcommand{\bQ}{{\overline Q}}
\newcommand{\bS}{{\overline S}}
\newcommand{\bD}{{\overline D}}
\newcommand{\hS}{{\hat S}}
\newcommand{\hbS}{{\hat{\overline S}}}
\newcommand{\bt}{{\bar\theta}}
\newcommand{\bpsi}{{\bar\psi}}
\newcommand{\bxi}{{\bar\xi}}
\newcommand{\p}[1]{(\ref{#1})}

\documentstyle[12pt]{article}

\begin{document}
\setcounter{page}0
\renewcommand{\thefootnote}{\fnsymbol{footnote}}
\thispagestyle{empty}
\begin{flushright}
hep-th/0210196
\vspace{2cm}\\ \end{flushright}
\begin{center}
{\large\bf Conformal and Superconformal Mechanics Revisited}
\vspace{0.8cm} \\
E. Ivanov,\footnote{eivanov@thsun1.jinr.ru }
S. Krivonos$\,$\footnote{krivonos@thsun1.jinr.ru }   \vspace{0.3cm} \\
{\it Bogoliubov  Laboratory of Theoretical Physics, JINR,\\
141980 Dubna, Moscow Region, Russia} \vspace{0.3cm} \\
and \vspace{0.3cm}\\
J. Niederle$\,$\footnote{niederle@fzu.cz}
  \vspace{0.3cm} \\
{\it Institute of Physics, Academy of Sciences of the Czech Republic, \\
Prague 8, CZ 182 21, Czech Republic}
\vspace{0.8cm}

{\bf Abstract}

\end{center}

\noindent We find, at the Lagrangian off-shell level, the explicit equivalence
transformation which relates the conformal
mechanics of De Alfaro, Fubini and Furlan to the conformal mechanics
describing the radial motion of the charged massive particle in the
Bertotti-Robinson AdS$_2\times S^2$ background. Thus we demonstrate the
classical equivalence of these two systems which are usually regarded as
essentially different ``old'' and ``new'' conformal mechanics models. We
also construct a similar transformation for $N=2$, $SU(1,1|1)$
superconformal mechanics in $N=2$ superfield formulation. Performing this 
transformation in
the action of the $N=2$ superconformal mechanics, we find an off-shell
superfield action of $N=2$ superextension of Bertotti-Robinson particle.
Such an action has not been given before. We show its
on-shell equivalence to the AdS$_2$ superparticle action derived from
the spontaneous partial breaking of $SU(1,1|1)$ superconformal symmetry
treated as the $N=2$ AdS$_2$ supersymmetry.

\newpage
\renewcommand{\thefootnote}{\arabic{footnote}}
\setcounter{footnote}0
\section{Introduction} 
Conformal mechanics \cite{dff} and
its superconformal extensions \cite{AP,fr}
were first introduced and studied as the simplest (one-dimensional) models
of (super)conformal field theory. Recently, there was a revival of interest in these models
in the context of the AdS/CFT correspondence \cite{mald,gkp,wit} and the physics of black 
holes. It was argued \cite{nscm} (see also \cite{3}-\cite{6}) that
the so-called ``relativistic'' generalizations of (super)conformal mechanics are candidate 
for the conformal field theory
dual to AdS$_2$ (super)gravity in the AdS$_2$/CFT$_{1}$ version of
the above correspondence. The simplest model of that kind is a particle evolving on the
 AdS$_2\times S^2$
background (the Bertotti-Robinson metric \cite{BR}) which describes the near-horizon 
geometry of $d=4$ Reissner-Nordstr\"om black hole.
The action (or Hamiltonian) of the standard (``non-relativistic'') conformal mechanics
is recovered from the worldline action (or Hamiltonian) of this model in the large 
black-hole mass approximation.

Both the ``old'' and ``new'' (super)conformal mechanics models respect the same 
(super)conformal symmetry,
though realized differently in either cases. This suggests that these models can in fact
be equivalent to each other.The aim of the present paper is to demonstrate,
on the simple examples of one-field conformal mechanics and its $N=2$ superextension, that 
such an equivalence indeed
takes place and is valid off-shell.\footnote{The issue of equivalence of various conformally
invariant $d=1$
models in the Hamiltonian formalism was addressed in \cite{Z}.}
We explicitly find the equivalence transformation of the time variable and involved fields 
which maps the ``old''
conformal mechanics actions on the actions of the AdS$_2$ (super)particle and vice versa. 
This off-shell equivalence holds at any finite
and non-zero value of the AdS$_2$ radius. Crucial for establishing this relation is the 
treatment of (super)conformal
mechanics models as a sort of $d=1$ sigma models on the cosets of $d=1$ conformal group 
$SL(2,R) \sim SO(1,2)$ and its superextensions \cite{ikl1,ikl2} (see also \cite{7}).
We demonstrate that the ``new'' (super)conformal models
correspond to another choice of these cosets as compared to the ``old'' ones and show how
to construct the relevant worldline actions in the spirit of the recent papers 
\cite{dik,bik3}.
Comparing the appropriate cosets, we establish the desired equivalence transformation
between two classes of (super)conformal mechanics models. In the superconformal case we
limit our consideration to the simplest example of $N=2$ superconformal 
mechanics \cite{AP,fr,ikl2},
but our arguments are valid also for higher $N$ cases treated in the nonlinear realizations 
approach of refs. \cite{ikl1,ikl2,7}. For the example considered we find
an off-shell $N=2$ superconformal action describing the radial motion of $N=2$ 
superextension
of the Bertotti-Robinson particle. Such an action has not been known before.
We show its on-shell equivalence to the AdS$_2$ superparticle action which
derives from the spontaneous partial breaking of $SU(1,1|1)$ superconformal
symmetry regarded as the $N=2$ AdS$_2$ supersymmetry.
\setcounter{equation}0
\section{Two versions of conformal mechanics}
We start by recalling basics of the bosonic case. The conformal
mechanics model of \cite{dff} is described by the following worldline action
\be
S = {1\over 2} \int dt \left( \mu \,\dot{x}^2 - {\gamma\over x^2} \right)~, \label{cmact}
\ee
which corresponds to the Hamiltonian
\be
H = {p^2\over 2 \mu} + {\gamma\over 2 x^2}~. \label{cmham}
\ee
The action \p{cmact} is invariant under the $d=1$ conformal transformations
\be
\delta t = a + b\, t + c\, t^2 \equiv a(t)~, \quad \delta x = {1\over 2}\,\dot{a}\,x =
{1\over 2} (b + 2c\,t)\,x  \label{conftr}
\ee
which form the $SO(1,2)$ group algebra:
\be\label{confbasis}
i \left[ P,D\right] =-P\; , \; i \left[ K,D\right] = K\; , \;i \left[ P,K \right] =-2D\; .
\ee

The generators $P, D$ and $K$ are, respectively, those of translations, dilatations and 
special
conformal transformations and correspond to the parameters $a, b$ and $c$ in \p{conftr}.
The Hamiltonian \p{cmham} is the translation operator in the canonical formalism.
It is straightforward to
construct the canonical charges corresponding to dilatations and conformal transformations.
It was argued in \cite{dff} that in the quantum case the true Hamiltonian is a combination 
of the above $H$ and the conformal generator, such that it generates a compact $SO(2)$ 
subgroup of $SO(1,2)$.

The ``new'' conformal mechanics model was proposed in \cite{nscm} as a model
of a charged massive particle
moving in the AdS$_2\times S^2$ Bertotti-Robinson (BR) background
\bea
ds^2 &= & -(2R/r)^4 d\tau^2 + (2R/ r)^2 dr^2 + R^2 d\Omega^2~, \nn
A &=& (2R/r)^2 d\tau~. \label{br}
\eea
Here, the parameter $R$ can be interpreted as the AdS$_2$ radius (equal to that of the 
sphere $S^2$),
$d\Omega^2 = d\theta^2 + \sin^2\theta d\varphi^2$ is the $SO(3)$ invariant metric on $S^2$ 
and $A$ is
the expression for the related Maxwell field in the coupled Einstein-Maxwell theory
to which the BR background \p{br} provides a solution. The static-gauge action of the 
particle and the corresponding canonical Hamiltonian are as follows
\bea
S &=& \int \left(qA - \tilde{\mu}\sqrt{-G(\tau)}\,d\tau\right) \nn
&=& \int d\tau (2R/r)^2 \left[q - \mu\,\sqrt{1 -(2R/r)^{-2}\dot{r}^2 -
R^2(2R/ r)^{-4} \left(\dot\theta^2 + \sin^2\theta \dot\varphi^2\right)}\right], 
        \label{snew} \\
H &=& (2R/r)^2 \left[ \sqrt{\mu^2 + (r^2p^2_r + 4L^2)/4R^2} - q \right]~, \label{hamnew} \\
L^2 &\equiv & p^2_\theta + \sin^{-2}\theta p^2_\varphi~.
\eea
The action \p{snew} is invariant under the field-dependent $d=1$ conformal transformations
\be
\delta \tau = a(\tau) + c (1/16R^2)\,r^4~, \quad \delta r = {1\over 2}\,\dot{a}\,r  ~, 
     \quad \delta \theta =
\delta \varphi = 0. \label{trantr}
\ee
As in the case of ``old'' conformal mechanics, in this case it is straightforward to find 
the canonical
Noether charges generating the ``active'' form of the conformal transformations  
\p{trantr} for the
involved fields \cite{nscm}.

In the limit $R \rightarrow \infty~, \;\tilde\mu - q \rightarrow 0~, \;R^2(q-\tilde\mu) 
\neq \infty$ the action \p{snew} and Hamiltonian \p{hamnew} go over into \p{cmact} and 
\p{cmham} with $x\equiv r$ and
\be
\mu = \tilde\mu~, \quad \gamma = 8R^2(\mu - q) + 4L^2/\mu~. \label{gamma1}
\ee
The ``new'' conformal mechanics was interpreted in \cite{nscm}-\cite{8} as 
a ``relativistic'' generalization of the ``old'' one just in this sense (the latter being 
a limiting case of the former).

The purely AdS$_2$ part of \p{snew}, \p{hamnew} corresponds to ``freezing'' the $S^2$ 
angles, $\theta = const, \varphi = const,
L^2 = 0$, and so describes the radial motion of the AdS$_2\times S^2$ particle.
In what follows we shall be interested just in
this reduced system which is in itself invariant under the conformal transformations
of $\tau$ and $r$ defined in \p{trantr}.
\setcounter{equation}0
\section{Conformal mechanics models as nonlinear realizations of $SO(1,2)$}
It was shown in \cite{ikl1} that both the action
\p{cmact} and equation of motion of the conformal mechanics of ref. \cite{dff} admit a 
transparent geometric interpretation in
terms of left-invariant Cartan 1-forms on the group $SO(1,2)$. Now we are going to show
that both the action \p{snew} (restricted to the radial motion) and non-standard conformal
transformations \p{trantr} can be also straightforwardly recovered from the same coset 
space approach.

We start by recalling some basic points of ref. \cite{ikl1}.

One proceeds from a nonlinear realization of the $SO(1,2)$ group acting as left shifts on 
the element
\be\label{gconf}
g=e^{itP} e^{iu(t)D} e^{i\lambda(t) K}~,
\ee
where $u(t)$ and $\lambda(t)$ are the Goldstone fields for the dilatation and special 
conformal
generators. The $SO(1,2)$ left shifts induce for $t, q(t)$ and $\lambda(t)$ the following
transformations
\be
\delta t = a +b\,t + c\,t^2 ~, \quad \delta u = b + 2c\,t ~, \delta \lambda = c\,e^u ~. 
 \label{nel1tran}
\ee
Next one defines the Cartan forms
\bea\label{formsconf}
&& g^{-1}d g = i\omega_P\,P +i\omega_D\,D + i\omega_K\,K~, \nn
&& \omega_P = e^{-u}dt~, \; \quad \omega_D = du-2 e^{-u} \lambda dt \; , \nn
&& \omega_K=d\lambda + e^{-u}\lambda^2 dt -\lambda du \;,
\eea
which by construction are invariants of the transformations \p{nel1tran}.
Using this fact, the coset field $\lambda(t)$ can be covariantly eliminated
by imposing the inverse Higgs \cite{IH} constraint
\be\label{IHconf}
\omega_D=0 \; \Rightarrow \; \lambda =\frac{1}{2}e^u \dot{u}
\;.
\ee
The manifestly invariant worldline action
\bea
S &=& -{1\over 2} \int \left(\mu \, \nu{}^2\,\omega_k + \gamma\, \nu{}^{-2} \,
     \omega_P\right) \nn
&=& {1\over 2} \int dt\left({1\over 4}\,\mu\,\nu{}^2\,e^u\, {\dot{u}}^2
-\gamma\,\nu{}^{-2}\,e^{-u} \right)~, \label{nel1act}
\eea
where $\nu$ $([\nu] = -1)$ is a normalization constant, is just the ``old'' conformal 
mechanics action \p{cmact} upon the identification
\be
x(t) = \nu\,e^{1/2\, u(t)}~.
\ee
The transformations \p{nel1tran}, being rewritten through $x(t)$, coincide
with \p{conftr}. Note that one can use in  \p{nel1act} the 1-forms with $u(t)$ and 
$\lambda(t)$ as independent
fields. Then the constraint \p{IHconf} arises as the equation of motion for $\lambda$.

The equation of motion for $x(t)$ or $u(t)$ following from \p{cmact} or \p{nel1act}
was interpreted in \cite{ikl1} as the equation for geodesics on $SO(1,2)$.
They can be time-like, space-like or null, depending on the choice
of the parameters $\gamma$ and $\mu$.

The basis \p{confbasis} in the algebra $so(1,2)$ can naturally be called ``conformal''
as it implies the standard $d=1$ conformal transformations for the time $t$.
Now we pass to another basis in the same algebra
\be\label{adsgenerators}
\hK =mK -\frac{1}{m}P\;,\; \hD=mD \;,
\ee
with $m$ being a parameter of the dimension of mass. This choice will be referred to as 
the ``AdS basis'' for a reason soon to be made clear.

The conformal algebra \p{confbasis} in the AdS basis \p{adsgenerators}
reads
\be\label{adsbasis}
i \left[ P,\hD\right] =-mP\; , \; i \left[ \hK,\hD\right] =2P+ m\hK\; , \;i \left[ P,\hK 
 \right] =-2\hD\; .
\ee
An element of $SO(1,2)$ in the AdS basis is defined to be
\be\label{confcoset}
g=e^{iyP}e^{i\phi(y) \hD} e^{i\Omega(y) \hK} \; .
\ee

Now we are in a position to explain the motivation for the nomenclature ``AdS basis''. 
With the choice
\p{gconf}, the group parameters $t$ and $u(t)$ parameterize the coset of $SO(1,2)$
over conformal generator $K$. So they are $d=1$ analogs of the standard $d=4$
Minkowski space-time 4-coordinate and dilaton which parameterize the coset of
the $d=4$ conformal group $SO(2,4)$ over the semi-direct product of
Lorentz group $SO(1,3)$ and the commuting abelian subgroup spanned by
generators of $d=4$ conformal boosts $K_m$. At the same time, the generator
$\hK$ \p{adsgenerators} can be shown to correspond to an $SO(1,1)$
subgroup of $SO(1,2)$. Thus the parameters $y$ and $\phi(y)$ in \p{confcoset}
parameterize the coset $SO(1,2)/SO(1,1)$, i.e. AdS$_2$. The parameterization
\p{confcoset} of AdS$_2$ is a particular case of the so-called ``solvable
subgroup parameterization'' of the AdS spaces \cite{solv}. The $d=4$ analog
of this parameterization is the parameterization of the AdS$_5$ space
in such a way that its coordinates are still parameters associated with
the 4-translation and dilatation generators $P_m, D$ of $SO(2,4)$, while it
is the subgroup $SO(1,4)$ with the algebra $ \propto \{P_m - K_m, so(1,3)\}$
which is chosen as the stability subgroup \cite{solv}.

The difference in the geometric meanings of the coordinate pairs $\left(t,
u(t)\right)$ and $\left(y, \phi(y)\right)$ is manifested in their different transformation
properties under the same $d=1$ conformal transformations. Left shifts of
the $SO(1,2)$ group element in the parameterization \p{confcoset} induce the
following transformations
\be
\delta y = a(y) + \frac{1}{m{}^2}\,c\,e^{2m\phi}~, \quad \delta \phi = \frac{1}{m}\,
\dot{a}(y) = \frac{1}{m}\,(b+2c\,y)~, \quad \delta \Omega = \frac{1}{m}\,c\,e^{m\phi} ~.
\label{modphitr}
\ee
We observe the modification of the special conformal transformation of $y$ by
a field-dependent term, just as in eq. \p{trantr}.

The relevant left-invariant Cartan forms are given by the following expressions
\bea\label{confcartan}
&& \hat\omega_D= \frac{1+\Lambda^2}{1-\Lambda^2} d\phi -2 \frac{\Lambda}{1-\Lambda^2} 
 e^{-m\phi}dy~, \nn
&& \hat\omega_P =\frac{1+\Lambda^2}{1-\Lambda^2} e^{-m\phi} d y - 2 
 \frac{\Lambda}{1-\Lambda^2} d\phi~, \nn
&& \hat\omega_K = m\frac{\Lambda}{1-\Lambda^2}\left( \Lambda e^{-m\phi} dy - d\phi\right) +
\frac{d\Lambda}{1-\Lambda^2}\,~,
\eea
where
\be\label{deflabda}
\Lambda = \tanh \Omega \;.
\ee

As in the previous realization, the field $\Lambda(y)$ can be eliminated by imposing the 
inverse Higgs constraint
\be\label{confih}
\hat\omega_D=0 \; \Rightarrow \; \partial_y \phi = 2\,e^{-m\phi}\,
 \frac{\Lambda}{1+\Lambda^2} \;,
\ee
whence $\Lambda $ is expressed in terms of $\phi $:
\be
\Lambda = \partial_y\phi\,e^{m\phi}\,\frac{1}{1 + \sqrt{1 - e^{2m\phi}\,
 (\partial_y\phi)^2}}~. \label{Lamb}
\ee

The $SO(1,2)$ invariant distance on AdS$_2$ can be defined, prior to imposing any 
constraints, as
\be
ds^2 = -\hat\omega_P^2 + \hat\omega_D^2 = -e^{-2m\phi}\,dy^2 + d\phi^2~.
\label{dist1}
\ee
Making the redefinition
$$
U = e^{-m\phi}~,
$$
it can be cast into the standard BR form
\be
ds^2 = -U^2\, dy^2 + (1/m^2)U^{-2}dU^2~, \label{dist2}
\ee
with $1/m$ as the inverse AdS$_2$ radius,
\be
{1\over m} = R~.
\ee
One more change of variable
\be
U = (2R/r)^2~, \quad y \equiv \tau
\ee
brings the distance just to the form \p{br} (with the $S^2$ part neglected).
The modified conformal transformations \p{modphitr} become just \p{trantr} after these
redefinitions.

The action \p{snew} can now be easily constructed from the Cartan forms \p{confcartan}
which, after substituting the inverse Higgs expression for $\Lambda $, eq. \p{Lamb},
read
\bea
&& \hat\omega_P = e^{-m\phi}\,\sqrt{1 - e^{2m\phi}\,(\partial_y\phi)^2 }\;dy~, \nn
&& \hat\omega_K = -\frac{m}{2}\,e^{-m\phi}\left(1 - \sqrt{1 - e^{2m\phi}\,(\partial_y\phi)^2
}\right)dy
+ \mbox{Total derivative}\times dy~. \label{invHforms}
\eea
The invariant action reads
\be
S = -\int \left(\tilde\mu \,\hat\omega_P - qe^{-m\phi} \right) =
-\int dy\, e^{-m\phi}\left(\tilde\mu\, \sqrt{1 -  e^{2m\phi}\,(\partial_y\phi)^2}
- q\right)~. \label{invact2}
\ee
After the above field redefinitions it is recognized as the radial-motion part of the 
``new'' conformal mechanics action \p{snew}. Note that the second term in \p{invact2}
is invariant under \p{modphitr} up to a total derivative in the integrand. The
action can be rewritten in a manifestly invariant form (with a tensor Lagrangian) by using
the explicit expression for $\hat\omega_K$ in \p{invHforms}
\be
S = \int \left[ (q-\tilde\mu)\,\hat\omega_P - (2/m)q\,\hat\omega_K \right]~.
\label{stens}
\ee
Note that, like in the previous case, in \p{stens} one can
use the Cartan forms with $\Lambda(y)$ as an independent field.
Then the inverse Higgs expression \p{Lamb} can be reproduced as the equation
of motion for $\Lambda(y)$ (with $\tilde\mu \neq 0$).

Now we are approaching the major point. We see that the ``old'' and ``new''
conformal mechanics models are associated with two different nonlinear
realizations of the same $d=1$ conformal group $SO(1,2)$ corresponding, respectively,
to the two different  choices \p{gconf} and \p{confcoset} of the parameterization of the 
group element.
The invariant actions in both cases can be written as integrals of linear combinations
of the left-invariant Cartan forms. But the latter {\it cannot} depend on the
choice of parameterization. Then the actions \p{nel1act} and \p{stens}
should in fact coincide with each other up to a redefinition of the free parameters of the 
actions. Thus
two conformal mechanics models are {\it equivalent} modulo redefinition of the
involved time coordinate and field. This statement should be contrasted with the
previous view of the ``old'' conformal mechanics model as a ``non-relativistic''
approximation of the ``new'' one.

To find the relation between actions \p{nel1act} and action \p{stens}, we first
write the relations between Cartan forms in two bases,
\be
\omega_K = m\,\hat\omega_K~, \quad \omega_P = \hat\omega_P - {1\over
m}\,\hat\omega_K~,
\ee
which follow from the definition \p{adsgenerators}. Then we
substitute these relations into \p{nel1act} and compare the resulting action
of the ``old'' conformal mechanics (thus rewritten in the parameterization
\p{confcoset}) with the action \p{stens}. We find that the two actions
coincide with each other, provided that their parameters are related as
\be
\gamma = 2\nu{}^2(\tilde\mu - q)~, \quad \mu = 2(R^2/\nu^2)(\tilde\mu + q)~.
\ee
Taking into account that \p{nel1act} is invariant under rescalings
$$
\gamma \rightarrow \gamma\,l~, \quad \mu \rightarrow \mu\,l^{-1}~, \quad \nu
\rightarrow \nu\,l^{1/2}~,
$$
one can choose, without loss of generality, the ``gauge''
\be
\nu^2 = 4R^2~, \label{gauge0}
\ee
in which
\be
\gamma = 8R^2(\tilde\mu - q)~, \quad \mu = {1\over 2}(\tilde\mu + q)~.
\label{rel12}
\ee

The same result can be obtained by directly performing in \p{cmact} or
\p{nel1act} the change of variables relating the conformal and AdS bases of
the $d=1$ conformal group. This transformation can be found by comparing the
parameterizations \p{gconf} and \p{confcoset}, with making use of the identity
\be
e^{it\hK} = e^{-\frac{i}{m}(\tanh t)\,P}e^{-2i (\ln \cosh t)\,D}e^{im (\tanh
t)\,K} \;,
\ee
which can be checked using \p{adsgenerators}, \p{adsbasis}. The sought
change of coordinates is as follows
\be
t=y-\frac{1}{m}\,e^{m\phi}\Lambda\;,\; u=m\,\phi +\ln \,(1-\Lambda^2)\; ,\; \lambda=m\,
 \Lambda \;. \label{conn}
\ee
These relations, together with \p{rel12} and the inverse Higgs expressions \p{IHconf}, 
 \p{Lamb} substituted for
$\lambda$, $\Lambda$, fully fix the equivalence map between the ``old'' and ``new'' 
conformal mechanics.

Finally, two remarks are in order.

First, for deriving the map \p{conn} it was essential that in our way of constructing
the AdS$_2$ particle action in terms of Cartan one-forms we started not just from
the coset $SO(1,2)/SO(1,1) \sim$ AdS$_2$,
but from the whole $SO(1,2)$ group space, including the parameter $\Lambda(y)$ associated 
with
the $SO(1,1)$ generator $\hat K$. This parameter has been eventually covariantly traded 
for $\phi, \partial_y\phi$
according to \p{Lamb}, but just its presence allowed us to equate the elements of
$SO(1,2)$ in the parameterizations \p{gconf}, \p{confcoset} and to derive the relations
\p{conn}. It would be difficult to guess these relations, while constructing the
AdS$_2$ particle action in the conventional way, just proceeding from a $d=1$
pullback of the AdS$_2$ metric \p{dist1}.

Secondly, the above map is well-defined at $R\neq \infty~, \;R\neq 0$, i.e.
it is one-to-one only for non-zero and finite value of the AdS$_2$ radius. The
action \p{snew} is still well defined at $m=0\, (R = \infty)$, in this limit it
becomes the static-gauge action of the massive particle in $d=2$ Minkowsky space.
However, such an action does not respect $d=1$ conformal symmetry,
its only symmetries are the $d=2$ translation and $SO(1,1)$ rotation ones.
Thus any link with the conformal mechanics turns out to be lost in this limit.
\setcounter{equation}0
\section{N=2 superconformal mechanics}
The previous consideration may be generalized to the supersymmetry case.
As the simplest example we consider in this Section $N=2$ superconformal mechanics
\cite{AP,fr,ikl2}. Like its bosonic prototype, $N=2$ superconformal mechanics has a natural 
description within the coset approach \cite{ikl2}.

The starting point is the $su(1,1|1)$ superalgebra which includes, apart from the $so(1,2)$ 
generators \p{confbasis}, those of Poincar\`{e} $\left\{ Q,\bQ \right\}$ and conformal
$\left\{ S,\bS \right\}$ supersymmetries and the $U(1)$ generator $U$. In the
conformal basis the non-vanishing (anti)commutators read:
\bea\label{superalgconf}
&& \left\{ Q,\bQ\right\}=-2P\;,\;\left\{ Q,\bS\right\}=-2D+2U\;,\;
\left\{ S,\bS\right\}=-2K\;,\;\left\{ S,\bQ\right\}=-2D-2U\;,\nn
&& i\left[ P, \left( \begin{array}{c} S \\ \bS \end{array} \right) \right]= -
   \left( \begin{array}{c} Q \\ \bQ \end{array} \right) \; , \;
   i\left[ K, \left( \begin{array}{c} Q \\ \bQ \end{array} \right) \right]=
   \left( \begin{array}{c} S \\ \bS \end{array} \right) \; , \nn
&& i\left[ D, \left( \begin{array}{c} Q \\ \bQ \end{array} \right) \right]= \frac{1}{2}
   \left( \begin{array}{c} Q \\ \bQ \end{array} \right) \; , \;
   i\left[ D, \left( \begin{array}{c} S \\ \bS \end{array} \right) \right]= -\frac{1}{2}
   \left( \begin{array}{c} S \\ \bS \end{array} \right) \; , \nn
&& i\left[ U, \left( \begin{array}{c} Q \\ \bQ \end{array} \right) \right]= \frac{1}{2}
   \left( \begin{array}{r} Q \\ -\bQ \end{array} \right) \; , \;
   i\left[ U, \left( \begin{array}{c} S \\ \bS \end{array} \right) \right]= \frac{1}{2}
   \left( \begin{array}{r} S \\ -\bS \end{array} \right) \; .
\eea
The standard realization of $SU(1,1|1)$ as the spontaneously broken $d=1, N=2$ 
superconformal group is set up by left multiplications of the coset
\be\label{superGconf}
g=e^{itP}e^{\theta Q+\bt\bQ}e^{iqD}e^{i\lambda K}e^{\psi S+ \bpsi\bS} \;,
\ee
where $(t, \theta, \bar\theta)$ are coordinates of $d=1, N=2$ superspace and
the remaining coset parameters are superfields given on this superspace.
The Poincar\`{e} supersymmetry transformations have the form
\be\label{stsusy}
\delta t= -i\left( \varepsilon \bt +{\bar\varepsilon} \theta\right)\;, \quad
\delta \theta = \varepsilon\;,\quad \delta\bt ={\bar\varepsilon} \;,
\ee
while the superconformal ones are given by
\be\label{confsusy}
\delta t= -it\left( \epsilon \bt +{\bar\epsilon} \theta\right), \;
\delta \theta = \epsilon\left( t+i\theta\bt\right),\;
\delta\bt ={\bar\epsilon}\left( t-i\theta\bt\right) ,\; \delta q=
-2i\left( \epsilon \bt +{\bar\epsilon} \theta\right)\;.
\ee

To single out the minimal set of coordinates and Goldstone superfields,
as well as to construct manifestly covariant dynamical equations of motion
for the latter, we first define the Cartan 1-forms
\bea\label{superformsconf}
&& \omega_P=e^{-q}d{\tilde t}\;,\;
\omega_D=dq-2e^{-q}\lambda d{\tilde t} +2ie^{-\frac{1}{2}q} \left( \psi d\bt+\bpsi d\theta 
    \right)\;, \nn
&& \omega_K=d\lambda +\lambda^2 e^{-q}d{\tilde t}-\lambda d q-
 2ie^{-\frac{1}{2}q}\lambda\left(\psi d\bt+
  \bpsi d\theta\right)+i\left( \psi d\bpsi+\psi d \bpsi \right) \;, \nn
&& \omega_Q=e^{-\frac{1}{2}q}d\theta - e^{-q}d{\tilde t}\psi\;, \;
\omega_S= d\psi-\frac{1}{2}\psi\left( dq-2e^{-q}\lambda d{\tilde t}\right)-
 e^{-\frac{1}{2}q}\left( \lambda+i\psi\bpsi\right)d\theta \;, \nn
&& \omega_{\bQ}=e^{-\frac{1}{2}q}d\bt - e^{-q}d{\tilde t}\bpsi\;, \;
\omega_{\bS}= d\bpsi-\frac{1}{2}\bpsi\left( dq-2e^{-q}\lambda d{\tilde t}\right)-
 e^{-\frac{1}{2}q}\left( \lambda-i\psi\bpsi\right)d\bt \;,
\eea
where
$$ d{\tilde t} = dt +i\left( \theta d\bt+\bt d\theta \right) \;.$$
Then, following ref. \cite{ikl2}, we subject them to the appropriate covariant
constraints.

The first kind of constraints is the inverse Higgs ones \cite{IH}:
\be\label{superIHconf}
\omega_D=0 \; \Rightarrow\; \left\{ \begin{array}{l}
 \dot{q}=2 e^{-q}\lambda \;, \\
 D q = 2ie^{-\frac{1}{2}q}\bpsi \;, \\
 \bD q = 2ie^{-\frac{1}{2}q} \psi\;, \end{array} \right.
\ee
 where
\be
D= \frac{\partial}{\partial\theta}+i\bt \partial_t\;,\; \bD=
\frac{\partial}{\partial\bt}+i\theta \partial_t\;,\; \left\{ D,\bD\right\}
=2i \partial_t \;.
\ee
As in the bosonic case, they are purely kinematical and
serve to covariantly express the redundant Goldstone superfields
$\lambda,\psi,\bpsi$ as derivatives of the only essential Goldstone
superfield, the dilaton $q(t,\theta,\bar\theta)$.

The equations of motion are produced by the extra covariant condition \cite{ikl2}
\be
\sqrt{\mu}\,\nu^2\,\omega_S =-i\sqrt{\gamma}\, \omega_Q~,
\ee
and they read
\be
\left[ D, \bD \right] Y= \frac{2}{\nu^2}\,\sqrt{\frac{\gamma}{\mu}}\,\frac{1}{Y} \;, \quad 
 Y=e^{\frac{1}{2}q} \;.
\ee

In contrast to the bosonic case, the lagrangian density of the action of $N=2$ 
superconformal mechanics
cannot be directly constructed out of the Cartan forms: it is not strictly invariant under 
superconformal
transformations and in this respect resembles the Wess-Zumino-Witten or Chern-Simons type 
Lagrangians. It has the following form
\be\label{superactionconf}
S_{conf}=\int dt d^2\theta \left( \frac{\mu \nu^2}{2}D Y \bD Y +
 \sqrt{\mu\gamma}\; \ln \,Y\right) \;,
\ee
and one can check that the Lagrangian in it is shifted by a total derivative under 
\p{confsusy}.
Such behavior of the Lagrangian is a key feature of most supersymmetric systems with
partial breaking of global supersymmetries (see e.g. \cite{A}).
This property makes the construction of the invariant minimal superfield actions in
superconformal mechanics a rather complicated problem as compared to the pure bosonic case.

The first component of superfield $q$ is just $u(t)\equiv q(t, \theta)|_{\theta =0}$ 
and the bosonic part of the action \p{superactionconf} coincides with the bosonic 
action \p{nel1act} (after eliminating the auxiliary field $[D,\bD]q|_{\theta = 0}$ 
by its equation of motion).

Now we shall consider a supersymmetric extension of the AdS basis
\p{adsgenerators}. The only new thing we have to do for this is to make
a rescaling of the superconformal generators as $\hS=mS, \hbS=m\bS$.
The superconformal algebra $su(1,1|1)$ in the AdS basis then reads:
\bea\label{superalgads}
&& \left\{ Q,\hbS\right\}=-2\hD+2mU\;,\; \left\{
\hS,\hbS\right\}=-2m\hK-2P\;,\;\left\{ \hS,\bQ\right\}=-2\hD-2mU\;,\nn &&
i\left[ P, \left( \begin{array}{c} \hS \\ \hbS \end{array} \right) \right]=
-m
   \left( \begin{array}{c} Q \\ \bQ \end{array} \right) \; , \;
   i\left[ \hK, \left( \begin{array}{c} Q \\ \bQ \end{array} \right) \right]=
   \left( \begin{array}{c} \hS \\ \hbS \end{array} \right) \; , \;
   i\left[ \hK, \left( \begin{array}{c} \hS \\ \hbS \end{array} \right) \right]=
   \left( \begin{array}{c} Q \\ \bQ \end{array} \right) , \nn
&& i\left[ \hD, \left( \begin{array}{c} Q \\ \bQ \end{array} \right) \right]= \frac{m}{2}
   \left( \begin{array}{c} Q \\ \bQ \end{array} \right) \; , \;
   i\left[ \hD, \left( \begin{array}{c} \hS \\ \hbS \end{array} \right) \right]= -
 \frac{m}{2}
   \left( \begin{array}{c} \hS \\ \hbS \end{array} \right) \; , \nn
&& i\left[ U, \left( \begin{array}{c} \hS \\ \hbS \end{array} \right) \right]= \frac{1}{2}
   \left( \begin{array}{r} \hS \\ -\hbS \end{array} \right) \; ,\;
i\left[ U, \left( \begin{array}{c} Q \\ \bQ \end{array} \right) \right]= \frac{1}{2}
   \left( \begin{array}{r} Q \\ -\bQ \end{array} \right) \; .
\eea

We define the realization of $SU(1,1|1)$ in the AdS basis by its left
action on the coset $SU(1,1|1)/U(1)$ in the following parameterization:
\be\label{superGads}
g=e^{iyP}e^{\theta Q+\bt\bQ}e^{i\Phi\hD}e^{i\Omega \hK}e^{\xi \hS+ \bxi\hbS} \;.
\ee

Left $SU(1,1|1)$ shift of the coset \p{superGads} induce the following superconformal
transformations for the coset parameters
\bea\label{superadstr}
&& \delta y =-i\left(\epsilon\bt+\bar\epsilon \theta\right)\left[ my+
 e^{m\Phi}\Lambda\right]-
i\left(\epsilon\bxi+\bar\epsilon \xi\right)e^{\frac{3}{2}m\Phi}\frac{1+\Lambda^2}{\sqrt{1-
 \Lambda^2}}\;, \nn
&& \delta \theta = m\epsilon\left( y+i\theta\bt -\frac{1}{m}e^{m\Phi}\Lambda\right)\;,\;
\delta \bt = m\bar\epsilon\left( y-i\theta\bt -\frac{1}{m}e^{m\Phi}\Lambda\right)\;, \nn
&& \delta \Phi =-2i\left(\epsilon\bt+\bar\epsilon \theta\right)-
2i\left(\epsilon\bxi+\bar\epsilon \xi\right)e^{\frac{1}{2}m\Phi}\frac{\Lambda}{\sqrt{1-
 \Lambda^2}}\;, \nn
&& \delta \xi=\epsilon e^{\frac{1}{2}m\Phi}\sqrt{1-\Lambda^2}+im\left(\bar\epsilon \theta-
 \epsilon\bt\right)\xi\;,\;
\delta \bxi=\bar\epsilon e^{\frac{1}{2}m\Phi}\sqrt{1-\Lambda^2}-im\left(\bar\epsilon \theta-
 \epsilon\bt\right)\bxi\;,\nn
&& \delta \Lambda =-im\left(\bar\epsilon \theta+\epsilon\bt\right)e^{\frac{1}{2}m\Phi}
 \sqrt{1-\Lambda^2} \;,
\eea
where $\Lambda$ is still defined according to \p{deflabda} as $\Lambda = \tanh \Omega$.

The relevant Cartan form needed to covariantly trade the superfields $\Omega,\xi,\bxi$ for
derivatives of $\Phi$ reads:
\be
\omega_{\hD}= \cosh (2\Omega)\,d\Phi - \sinh (2\Omega)\,e^{-m\Phi}d{\tilde y}+
 2i\cosh (\Omega)\,e^{-\frac{m}{2}\Phi}
 \left(\xi d\bt+\bxi d\theta \right) \;,
\ee
where
$$
d{\tilde y}=dy +i\left( \theta d\bt+\bt d\theta \right) \;.
$$
Putting this form equal to zero, we obtain
\be
\partial_y \Phi=2e^{-m\Phi}\frac{\Lambda}{1+\Lambda^2}\;, \;
D_y\Phi=2ie^{-\frac{m}{2}\Phi}\, \frac{\sqrt{1-\Lambda^2}}{1+\Lambda^2}\,\bxi\;, \;
\bD_y\Phi=2ie^{-\frac{m}{2}\Phi}\,
\frac{\sqrt{1-\Lambda^2}}{1+\Lambda^2}\,\xi\;,\label{IHexpr2}
\ee
where
\be
D_y= \frac{\partial}{\partial\theta}+i\bt \partial_y\;,\;
\bD_y= \frac{\partial}{\partial\bt}+i\theta \partial_y\;,\;
\left\{ D_y,\bD_y\right\} =2i \partial_y \;.
\ee

Now we are able to establish a link between the above two nonlinear
realizations. This equivalence transformation is a supersymmetric extension of \p{conn}.
The relations between the coset coordinates can be found by rearranging the exponents
{}from the parameterization \p{superGconf} to \p{superGads}. The explicit form
of these relations is as follows
\be\label{connsusy}
t=y-\frac{1}{m}\,e^{m\Phi}\Lambda\;,\; q=m\Phi + \ln \,(1-\Lambda^2)\; ,\; \lambda=m\Lambda
  \;,\; \psi=m\xi\;,\; \bpsi=m\bxi\;.
\ee

Now we have all the necessary ingredients for passing from one basis another
in the $N=2$ superconformally invariant actions. As the suggestive example,
let us rewrite the standard $N=2$ superconformal mechanics action \p{superactionconf}
in the AdS basis
\bea\label{adsaction0}
&& S= \frac{1}{2} \int dt d^2\theta \left( -\mu \nu^2\, \bpsi\psi + \sqrt{\mu\gamma}\, 
 q\right) = \frac{1}{2} \int dy d^2\theta
\left[ \frac{1-\Lambda^2}{1+\Lambda^2} -\frac{1}{m}e^{m\Phi}\partial_y \Lambda\right] \nn
&& \times \,\left[ \frac{1}{4} \mu \nu^2 m^2\,e^{m\Phi}\frac{ (1+\Lambda^2)^2}{1-\Lambda^2}
 D_y \Phi \bD_y \Phi+
\sqrt{\mu\gamma}\,\left( m\Phi + \ln(1-\Lambda^2)\right) \right].
\eea
Here
\be
\Lambda = e^{m\Phi}\, \partial_y\Phi\,\frac{1}{1+\sqrt{1-e^{2m\Phi}(\partial_y
\Phi)^2}}~, \;\; \Lambda^2 = \frac{1 - \sqrt{1-e^{2m\Phi}(\partial_y
\Phi)^2}}{1 +\sqrt{1-e^{2m\Phi}(\partial_y
\Phi)^2}}
\ee
as follows from \p{IHexpr2}. From now on we shall use the ``gauge''
\p{gauge0}:
\be
\nu^2m^2 = 4 R^2 m^2 = 4\,. \label{gaug2}
\ee

While neglecting the superpotential term (i.e. choosing $\gamma=0$) and terms
with higher derivatives (those with $\partial_y \Lambda$),  the action is
simplified to the form
\be\label{superadsaction}
S = \mu\,\int dyd^2\theta \left[ \frac{e^{m\Phi} D_y \Phi \bD_y \Phi}{1+
  \sqrt{1-e^{2m\Phi}\,(\partial_y \Phi)^2}}  \right],
\ee
which is the product of the first terms in both expressions inside the square brackets in
\p{adsaction0}. One can directly check that the bosonic part of \p{superadsaction} just
coincides with \p{stens} (for $\tilde\mu = q$) after elimination of the
auxiliary field. Using the definition $d^2\theta = \bar D D$, and
suppressing the fermionic fields, one finds
\bea
S^0_{bos} &\equiv & \int dy \,L_0 = {\mu\over 2} \int dy \left[ (\partial_y\phi)^2 + F^2
\right]e^{m\phi} \left(1 + \Lambda^2\right) \nn
&=& \mu \int dy \left[ (\partial_y\phi)^2 + F^2
\right]e^{m\phi} \frac{1}{1 + \sqrt{1-e^{2m\phi}(\partial_y \phi)^2}}\,.
\label{S0bos}
\eea
On shell, i.e. with $F=0$, we obtain
\be\label{bosads}
S^0_{bos} = \mu \int dy e^{-m\phi} \left[1 -\sqrt{1-e^{2m\phi}(\partial_y
\phi)^2}\right]\,.
\ee

The full action \p{adsaction0} looks rather complicated and includes terms
which could seemingly give rise
to higher derivatives in $y$ in the component Lagrangian. Nevertheless, this does not 
happen. By direct calculation, the remaining contributions to the bosonic part of the
off-shell component action are given by
\be
S^1_{bos} = \int dy \left( L_1 + L_2 \right)~, \label{S11}
\ee
where
\bea
&& L_1 =  -\frac{\mu}{4m}\,\left[ (\partial_y\phi)^2 + F^2
\right]e^{3m\phi}\,\frac{(1 + \Lambda^2)^4}{(1 - \Lambda^2)^2}
\left[\partial^2_y \phi +m (\partial_y\phi)^2 \right], \label{long0} \\
&& L_2 = -{1\over 2}\,\sqrt{\mu\gamma}\,F \left\{m - e^{2m\phi}\,
\frac{(1+ \Lambda^2)^3}{2(1 - \Lambda^2)^2}\left[\partial^2_y \phi +m (\partial_y\phi)^2 
 \right] \right\}. \label{long}
\eea
After the change of variable
$$
X = {1\over m} e^{m\phi}\, \quad \Rightarrow \quad \Lambda \equiv \Lambda
(\partial_yX) = \partial_y X \,\frac{1}{1 + \sqrt{1 - (\partial_y X)^2}}\,,
$$
the $F$-independent term in \p{long0} can be shown to be a
full derivative
$$
\sim \partial_y^2 X \,f(\partial_y X)~.
$$
Thus \p{S11} contributes only to the potential of the auxiliary field $F(y)\,$.
The algebraic equation for the latter which follows from $S_{bos} = S^0_{bos} +
S^1_{bos}$ is easily found to be
\be
F = {m\over 4}\,\sqrt{\frac{\gamma}{\mu}}\,e^{-m\phi}\,\left[1 +
\sqrt{1-e^{2m\phi}(\partial_y\phi)^2}\right].
\ee
After substituting this back into $S_{bos}\,$, the latter, modulo a full
derivative in the Lagrangian, reads
\be
S_{bos} = \int dy e^{-m\phi}\left\{ \mu\left[1
-\sqrt{1-e^{2m\phi}(\partial_y \phi)^2}\right] -
\frac{\gamma}{16}\,m^2\,\left[1 +\sqrt{1-e^{2m\phi}(\partial_y
\phi)^2}\right]\right\},
\ee
which coincides with \p{invact2} under the identification \p{rel12}.

Thus \p{adsaction0} provides a manifestly $N=2$ supersymmetric off-shell form of
$N=2$ superconformal extension of the ``new'' conformal mechanics action \p{invact2}
which describes the radial (AdS$_2$) motion of the charged particle in the BR
AdS$_2\times S^2$ background. Such a superfield action was not known
before. By construction, it is related by the equivalence transformation
\p{connsusy} to the ``old'' $N=2$ superconformal mechanics action \p{superactionconf}.

As the last topic, we would like to show that the action \p{adsaction0} at $\gamma =0$ is 
equivalent,
at least  on-shell, to the ``standard'' action for the superparticle in AdS$_2$ background.
The latter action can be obtained by dimensional reduction from that of $N=1$ supermembrane 
in AdS$_4$
background \cite{dik,IK11} and has the following form:
\be\label{actionads}
\tilde{S}= \mu\int dyd^2\theta {\cal L} \equiv
  \mu\,\int dyd^2\theta \left[ \frac{e^{m\tilde{\Phi}} D_y \tilde{\Phi} \bD_y 
\tilde{\Phi}}{1+
  \sqrt{1+e^{2m\tilde{\Phi}}D_y\bD_y \tilde{\Phi} \bD_y D_y \tilde{\Phi}}}
\right], \quad \tilde{\Phi}|_{\theta=0} = \Phi|_{\theta =0}~.
\ee
This action can be shown to be invariant (up to a shift of the Lagrangian by a
full derivative) under the following transformations
\be \delta
\tilde{\Phi} = L \tilde{\Phi} -2i \left( \epsilon\bar\theta+\bar\epsilon\theta\right)
+ie^{m\tilde{\Phi}}
 \left( \epsilon D_y {\cal L} + \bar\epsilon \bD_y {\cal L}\right),
\ee
where
\be
L = im\epsilon\left( \theta\bar\theta +2iy\right) \frac{\partial}{\partial\theta} - 
 im\bar\epsilon
\left( \theta\bar\theta -2iy\right)\frac{\partial}{\partial\bar\theta} +
 2im\left(\bar\epsilon\theta+
 \epsilon\bar\theta\right)y\partial_y~.
\ee
These transformations, although being different from \p{superadstr}, still
constitute a nonlinear realization of the same $N=2$ superconformal group
$SU(1,1|1)$, with $\tilde{\Phi}$ being a Goldstone dilaton superfield.

The purely bosonic part of \p{actionads} reads
\be\label{tildeS} \tilde{S}_{bos}
= \mu \int dy \left[ (\partial_y\phi)^2 + \tilde{F}^2 \right]e^{m\phi}
\frac{1}{1 + \sqrt{1-e^{2m\phi}[(\partial_y \phi)^2 + \tilde{F}^2]}}
\ee
and it obviously coincides on shell, when $F = \tilde{F} =0$,
with the bosonic part of \p{adsaction0} at $\gamma = 0$, viz.
$S_{bos}(\gamma=0)  = \int dt \left(L_0 + L_1\right)\,$.\footnote{Actually, 
it is easy to find an invertible
relation between the auxiliary fields $F$ and $\tilde{F}$ to show that
$\tilde{S}_{bos}$ and $S_{bos}(\gamma = 0)$ are off-shell equivalent.} Thus
it remains to demonstrate that in both actions, \p{adsaction0} (at $\gamma =0$)
and \p{actionads},  the fermionic terms on shell also coincide with
each other modulo an equivalence redefinition of the fermionic fields.
The proof is based on the following common feature of both actions: after
elimination of the auxiliary fields $F$ and $\tilde{F}$ by their equations of
motion, the fermionic terms in \p{adsaction0} and \p{actionads} take the
following generic form
\be\label{fermiterms}
a_1 (\partial_y \psi \bar\psi - \psi\partial_y\bar\psi ) + a_2 \psi\partial_y \psi \bar\psi
 \partial_y \bar\psi \;.
\ee
Here,  $a_1$ and $a_2$ are some functions of $\phi$ and $\partial_y\phi$
which are specific for either actions. The crucial observation is that the invertible
redefinition of the fermionic variables
\be\label{redef1}
\psi\rightarrow \tilde\psi=\sqrt{a_1} \psi - \frac{a_2}{2\sqrt{a_1}} \psi\partial_y \psi 
 \bar\psi \;, \quad
\bar \psi\rightarrow \widetilde{\bar\psi}= \sqrt{a_1} \bar\psi - \frac{a_2}{2\sqrt{a_1}} 
 \psi\partial_y \bar\psi
\bar\psi  \;,
\ee
reduces fermionic terms in both actions to the pure kinetic term
\be
\partial_y \tilde\psi \widetilde{\bar\psi} - \tilde\psi \partial_y\widetilde{\bar\psi}
\ee
for arbitrary functions $a_1$ and $a_2$ (the function $a_1$ is assumed to start with a 
constant,
so it is legitimate to divide by it in \p{redef1}). Keeping in mind that the bosonic parts 
of the
on-shell actions are the same, we conclude that on-shell the actions \p{actionads} and 
\p{adsaction0} coincide with each other, modulo an equivalence redefinition of the fermionic
components. It would be instructive to find a superfield redefinition relating
$\Phi$ and $\tilde{\Phi}$ and to prove the off-shell equivalence of
\p{actionads} and \p{adsaction0}.
\setcounter{equation}0
\section{Conclusions} In this paper we have found an equivalence
transformation between two nonlinear realizations of the group $SO(1,2)$:
its realization as the spontaneously broken $d=1$ conformal symmetry and the
realization as the AdS$_2$ isometry group. This transformation takes the
action of the ``old'' conformal mechanics into the action of the ``new'' one, viz. the 
worldline action describing the radial motion of a charged BR particle.
Thus these systems, usually treated as two essentially different models of conformal 
mechanics, prove in fact to be classically equivalent. We extended this equivalence 
transformation to the case of $N=2$ superconformal mechanics. Being applied to the 
standard $N=2$ superfield 
action of the latter, this transformation produces a novel superfield action which 
provides an $N=2$ superextension of the ``new'' conformal mechanics action and so describes 
a radial motion of $N=2$ BR superparticle in a static gauge. Similar equivalence 
transformations  can hopefully be established for higher $N$ superconformal mechanics
 models in the nonlinear realization approach of refs.\cite{ikl1,ikl2,7}. 

It is very interesting to elaborate on the quantum implications of the
equivalence transformation found. We expect that it will allow one to explicitly
solve the quantum mechanics of the BR particle and its superextensions in
terms of (super)conformal quantum mechanics. Also, it would be worth
to extend our consideration to the full BR  AdS$_2\times S^2$ background
and its superextensions, thus taking into account the angular degrees of
freedom of the BR (super)particle. The corresponding ``old'' conformal mechanics 
action should include a conformally invariant $d=1$ $S^2$ sigma model part. It could
naturally appear as a bosonic core of the action of one of possible versions of $N=4$ 
superconformal mechanics, e.g. associated with the superconformal group $SU(1,1|2)$.

\section*{Acknowledgements} This work was partially fulfilled during a few collaborative
visits of E.A. and S.K. to Institute of Physics in Prague in the framework of the 
Votruba-Blokhintsev
Program. They thank the Directorate of the Institute for hospitality extended to them  and 
the Votruba-Blokhintsev Program for financial support. Their work was also partially 
supported by an INTAS grant, project No. 00-00254, DFG grant, project No. 436 RUS 113/669, 
RFBR-DFG grant, project No. 02-02-04002 and RFBR-CNRS grant, project No. 01-02-22005.

\end{document}